\documentclass[twoside,twocolumn,9pt]{article}
\usepackage{extsizes}
\usepackage[super,sort&compress,comma]{natbib} 
\usepackage[version=3]{mhchem}
\usepackage[left=1.5cm, right=1.5cm, top=1.785cm, bottom=2.0cm]{geometry}
\usepackage{balance}
\usepackage{times,mathptmx}
\usepackage{sectsty}
\usepackage{graphicx} 
\usepackage{lastpage}
\usepackage[format=plain,justification=justified,singlelinecheck=false,font={stretch=1.125,small,sf},labelfont=bf,labelsep=space]{caption}
\usepackage{float}
\usepackage{fancyhdr}
\usepackage{fnpos}
\usepackage[english]{babel}
\addto{\captionsenglish}{%
  
}
\usepackage{array}
\usepackage{droidsans}
\usepackage{charter}
\usepackage[T1]{fontenc}
\usepackage[usenames,dvipsnames]{xcolor}
\usepackage{setspace}
\usepackage[compact]{titlesec}
\usepackage{hyperref}

\usepackage{epstopdf}

\definecolor{cream}{RGB}{222,217,201}

\begin{document}

\pagestyle{fancy}
\thispagestyle{plain}
\fancypagestyle{plain}{

\fancyhead[C]{\includegraphics[width=18.5cm]{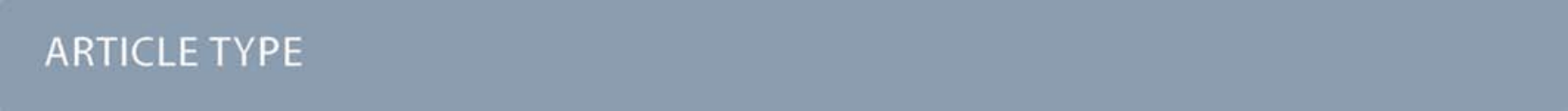}}
\fancyhead[L]{\hspace{0cm}\vspace{1.5cm}\includegraphics[height=30pt]{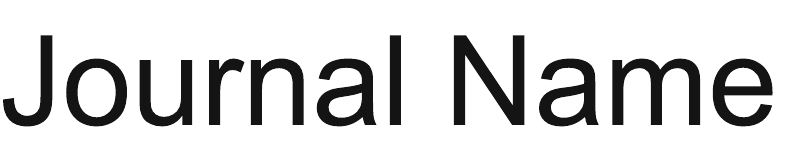}}
\fancyhead[R]{\hspace{0cm}\vspace{1.7cm}\includegraphics[height=55pt]{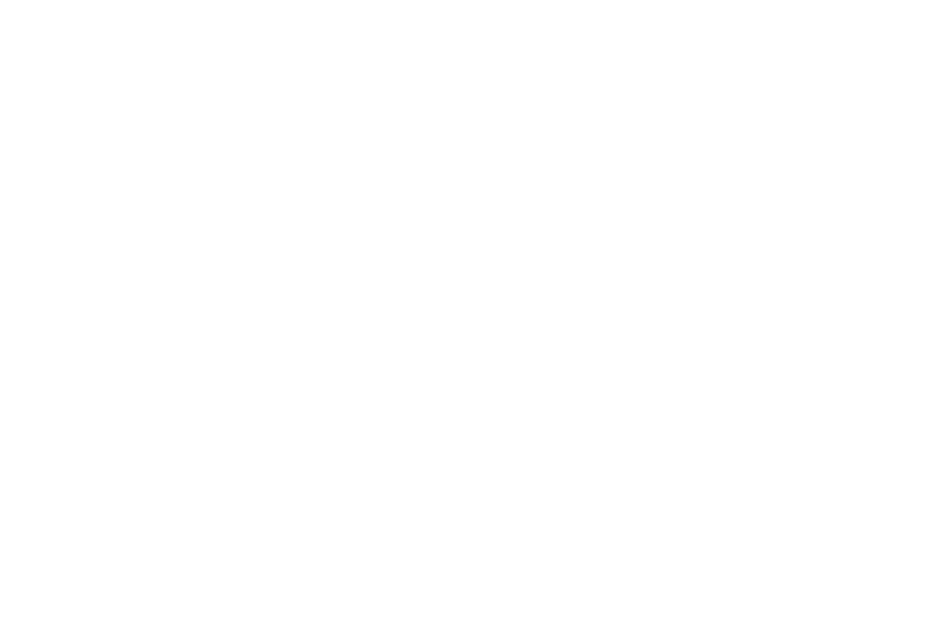}}
\renewcommand{\headrulewidth}{0pt}
}

\makeFNbottom
\makeatletter
\renewcommand\LARGE{\@setfontsize\LARGE{15pt}{17}}
\renewcommand\Large{\@setfontsize\Large{12pt}{14}}
\renewcommand\large{\@setfontsize\large{10pt}{12}}
\renewcommand\footnotesize{\@setfontsize\footnotesize{7pt}{10}}
\makeatother

\renewcommand{\thefootnote}{\fnsymbol{footnote}}
\renewcommand\footnoterule{\vspace*{1pt}%
\color{cream}\hrule width 3.5in height 0.4pt \color{black}\vspace*{5pt}} 
\setcounter{secnumdepth}{5}

\makeatletter 
\renewcommand\@biblabel[1]{#1}            
\renewcommand\@makefntext[1]%
{\noindent\makebox[0pt][r]{\@thefnmark\,}#1}
\makeatother 
\renewcommand{\figurename}{\small{Fig.}~}
\sectionfont{\sffamily\Large}
\subsectionfont{\normalsize}
\subsubsectionfont{\bf}
\setstretch{1.125} 
\setlength{\skip\footins}{0.8cm}
\setlength{\footnotesep}{0.25cm}
\setlength{\jot}{10pt}
\titlespacing*{\section}{0pt}{4pt}{4pt}
\titlespacing*{\subsection}{0pt}{15pt}{1pt}

\fancyfoot{}
\fancyfoot[LO,RE]{\vspace{-7.1pt}\includegraphics[height=9pt]{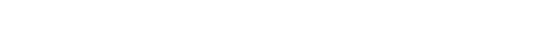}}
\fancyfoot[CO]{\vspace{-7.1pt}\hspace{13.2cm}\includegraphics{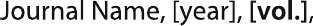}}
\fancyfoot[CE]{\vspace{-7.2pt}\hspace{-14.2cm}\includegraphics{head_foot/RF}}
\fancyfoot[RO]{\footnotesize{\sffamily{1--\pageref{LastPage} ~\textbar  \hspace{2pt}\thepage}}}
\fancyfoot[LE]{\footnotesize{\sffamily{\thepage~\textbar\hspace{3.45cm} 1--\pageref{LastPage}}}}
\fancyhead{}
\renewcommand{\headrulewidth}{0pt} 
\renewcommand{\footrulewidth}{0pt}
\setlength{\arrayrulewidth}{1pt}
\setlength{\columnsep}{6.5mm}
\setlength\bibsep{1pt}

\makeatletter 
\newlength{\figrulesep} 
\setlength{\figrulesep}{0.5\textfloatsep} 

\newcommand{\topfigrule}{\vspace*{-1pt}%
\noindent{\color{cream}\rule[-\figrulesep]{\columnwidth}{1.5pt}} }

\newcommand{\botfigrule}{\vspace*{-2pt}%
\noindent{\color{cream}\rule[\figrulesep]{\columnwidth}{1.5pt}} }

\newcommand{\dblfigrule}{\vspace*{-1pt}%
\noindent{\color{cream}\rule[-\figrulesep]{\textwidth}{1.5pt}} }

\makeatother

\twocolumn[
  \begin{@twocolumnfalse}
\vspace{3cm}
\sffamily
\begin{tabular}{m{4.5cm} p{13.5cm} }

\includegraphics{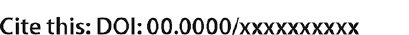} & \noindent\LARGE{\textbf{Directional Clogging and Phase Separation for Disk Flow Through Periodic and Diluted Obstacle Arrays}} \\
\vspace{0.3cm} & \vspace{0.3cm} \\

& \noindent\large{C. Reichhardt and C. J. O. Reichhardt} \\

\includegraphics{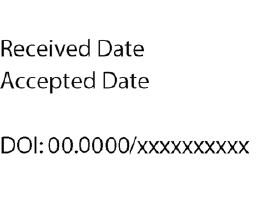} & \noindent\normalsize{
We model collective disk flow though a square array of obstacles as the flow direction is changed relative to the symmetry directions of the array. At lower disk densities there is no clogging for any driving direction, but as the disk density increases, the average disk velocity decreases and develops a drive angle dependence. For certain driving angles, the flow is reduced or drops to zero when the system forms a heterogeneous clogged state consisting of high density clogged regions coexisting with empty regions. The clogged states are fragile and can be unclogged by changing the driving angle.  For large obstacle sizes, we find a uniform clogged state that is distinct from the collective clogging regime. Within the clogged phases, depinning transitions can occur as a function of increasing driving force, with strongly intermittent motion appearing just above the depinning threshold.  The clogging is robust against the random removal or dilution of the obstacle sites, and the disks are able to form system-spanning clogged clusters even under increasing dilution. If the dilution becomes too large, however, the clogging behavior is lost.
} \\

\end{tabular}

 \end{@twocolumnfalse} \vspace{0.6cm}

  ]

\renewcommand*\rmdefault{bch}\normalfont\upshape
\rmfamily
\section*{}
\vspace{-1cm}


\footnotetext{\textit{Theoretical Division and Center for Nonlinear Studies, Los Alamos National Laboratory, Los Alamos, New Mexico 87545, USA. Tel: +1 505 665 1134; E-mail: cjrx@lanl.gov}}




There are a variety of systems that can be described as a loose assembly of
particles which exhibit jamming behaviors.
At lower densities, flow occurs easily in such systems,
but at high densities the system can act like a solid in which all flow
ceases
\cite{Liu98,OHern03,Drocco05,Liu10,Reichhardt14,Olsson07}.
Jamming has been extensively studied 
as a function of
density \cite{Liu98,Drocco05}, shear \cite{Olsson07},
particle shape \cite{Reichhardt14,Zou09,Lopatina11,Hoy17},
and friction effects \cite{Candelier10,Bi11}.
Many of these studies involved
no quenched disorder 
so that the system can be described
as containing only particle-particle interactions.
It is also possible for the
particle motion to be stopped by some form of external
constraints, such as
flow through bottlenecks or funnels
\cite{To01,Thomas13,Chen12,Zuriguel15,Gella18,Reichhardt18},
motion through a mesh \cite{Redner00,Chevoir07,Roussel07,Wyss06,Barre17},
flow over a disordered substrate 
\cite{Reichhardt12,Reichhardt18c,LeBlay20},
or flow in porous media
\cite{Alava04,Liu95,Wirner14,Dressaire17,Liu19,Gerber18,Gerber19}.
The particle flow stops when the combination of the particle density and
the obstacle density is high enough.
Open questions include whether the cessation of flow
in systems with
quenched disorder is best described as jamming or clogging, as well as
how to distinguish between these two phenomena.

There are several limiting
cases for jamming and clogging behavior.
For example, 
frictionless disks have a well defined jamming density $\phi_J$ in the
absence of obstacles.
If a small number of obstacles are added,
the system can still be described as reaching a jamming point at a slightly
lower density $\phi<\phi_J$ due to the diverging
length scale $l_{J}$ that emerges as the
jamming density in the clean system is approached.
Jamming, which is associated with a uniform particle density throughout
the system, occurs once
the average distance between obstacles $l_{\rm obs}$ becomes smaller than
the jamming length scale,
$l_{\rm obs} < l_{J}$.
Another limit is the clogging of a single particle, which
can arise for flow along the $x$ direction through a
square array of obstacles when the obstacle radius becomes large enough that
the particle cannot fit in the space between adjacent obstacles.
Between the jamming and single particle clogging limits,
a variety of other types of
collective clogging behaviors are possible in which
groups of particles come together to create a locally stuck region. 

Several studies addressing the effects of a small number of
obstacles or weak quenched disorder on the jamming transition show
that the jamming density decreases as obstacles are added
\cite{Brito13,Graves16,WentworthNice20}, while other studies have
focused on a crossover from jamming to clogging behavior
for particles moving through obstacle arrays
\cite{Nguyen17,Peter18,Stoop18,Leyva20}. 
P{\' e}ter {\it et al.} \cite{Peter18} considered
an assembly of monodisperse particles moving over a random 
obstacle array.
For a small number of obstacles, they found jamming behavior in which the
particle density is uniform in the motionless state.
Once the obstacle density exceeds a certain threshold, 
the system exhibits a clogged state down to very low particle densities.
The clogged state is highly 
heterogeneous
and contains local patches in which the particle density is close
to the jamming density along with other patches in which there are few
or no particles.
Additionally, the system requires time in which to 
gradually organize into a particular clogged configuration,
whereas the jammed states form very rapidly.
Nguyen {\it et al.} \cite{Nguyen17}
studied an assembly of bidisperse grains moving through a 
two-dimensional periodic obstacle array
and also found a transition to a clogged state characterized
by the formation of a high density phase coexisting with 
a low density phase.
In this case, the susceptibility to the formation of a clogged state
depended on the direction of the flow relative to the
substrate symmetry directions.
For example, when the obstacles are small,
the system does not jam when driven 
along the $x$-direction, but for driving along $30^\circ$,
the system can reach a clogged or partially jammed
state.
In experimental work, Stoop {\it et al.} \cite{Stoop18,Leyva20}
studied the motion of colloidal disks
through
a random array of obstacles.
They found that the flow decreases over time due to the gradual formation
of clogged regions. 

In this work we examine monodisperse disks
moving through a periodic array of obstacles under an external drive that
varies in direction from
$0^\circ$ to $90^\circ$ from the $x$ axis.
For low disk densities, the disks flow for every direction of
applied drive; however, the net velocity drops
at incommensurate angles 
and reaches a maximum for drives along
the easy flow directions of $0^\circ$, $45^\circ$ and $90^\circ$.
When the disk density is increased,
we find that although flow still occurs for driving near
$0^\circ$ and $90^\circ$, the
system begins to clog at the intermediate angles by forming
a phase separated state containing regions of high and low disk density.
The clogged system is fragile in nature \cite{Cates98}, and we find a
partial hysteresis effect in which the flow can resume if the driving
angle is changed after the clogged state forms.
We map the locations of the clogged states as a function of obstacle size,
and show that there is a critical obstacle size above which even
single disks become clogged for driving along the incommensurate directions.
We find that a clogged state can undergo
a depinning transition to a flowing state if the driving force is
increased.
Just above the depinning transition, the flow is strongly intermittent
and there is a coexistence of clogged states
and moving states; however, the moving disks do not exchange neighbors,
indicating
that the depinning transition is elastic \cite{Reichhardt17}. 
We find that the clogged states
are fairly robust to dilution of the obstacle lattice
as long as
large scale system-spanning dense clusters can still occur;
however, when the dilution becomes too strong,
the system flows instead of clogging.

Experimental systems in which our results could be tested
include particle flow through periodic obstacle arrays
\cite{Huang04,Li07,Long08,McGrath14,Wunsch16,Stoop20} 
or optical trap arrays
\cite{Korda02,MacDonald03,Bohlein12,Juniper16,Cao19}.
Most previous works in such systems were performed
in the low density
regime where particle-particle interactions are weak.
Clogging behavior is expected to occur for high particle densities
or in regimes where the diameter of the obstacles is large.

\subsection{Simulation}
We model 
a two dimensional $L \times L$ system
containing a square array of obstacles with lattice spacing $a$
and obstacle radius $r_{\rm obs}$.
We fix $L=36$ and $a=4.0$.
Within the system we place $N_{d}$ mobile disks
with dynamics given
by the following overdamped equation of motion: 
\begin{equation} 
\alpha_d {\bf v}_{i}  =
{\bf F}^{dd}_{i} +  {\bf F}^{obs}_{i} + {\bf F}^{D}_{i}.
\end{equation}
The velocity of disk $i$
is ${\bf v}_{i} = {d {\bf r}_{i}}/{dt}$, 
the disk position is ${\bf r}_{i}$, and the damping constant  
$\alpha_d$ is set to $\alpha_d=1.0$. 
The disk-disk interaction force 
${\bf F}^{dd}_{i}$ arises from a harmonic repulsive potential with
radius $r_{d}$, which we fix to $r_d=0.5$. 
The disk-obstacle force ${\bf F}^{obs}$ is also modeled with a repulsive 
harmonic potential.
The system density is defined as the
total area covered by the obstacles and mobile disks, 
$\phi = N_{\rm obs}\pi r^2_{\rm obs}/L^2 + N_{d}\pi r^2_{d}/L^2$,
where $N_{\rm obs}$ is the number of obstacles. 
The external drive
${\bf F}_{D} = F_{D}[\cos(\theta){\bf \hat{x}} + \sin(\theta){\bf \hat{y}}]$ 
is initially applied
along the $x$-direction and gradually rotates from
$\theta=0^\circ$ to $\theta=90^\circ$ or higher. 
We measure the average velocity of the disks in the $x$-direction, 
$\langle V_{x}\rangle = \langle \sum^{N_{d}}_{i= 1}{\bf v}_i\cdot {\bf \hat{x}}\rangle$, where the average is taken over time,
and in the $y$-direction,
$\langle V_{y}\rangle = \langle \sum^{N_{d}}_{i= 1}{\bf v}_i\cdot {\bf \hat{y}}\rangle$,
as well as the net velocity
$\langle V\rangle = \sqrt{\langle V_x\rangle^{2} + \langle V_y\rangle^{2}}$.
We find that the dynamics can depend on the
rate at which the drive direction is changed,
so we consider the limit where the
direction is changed slowly enough that 
such effects are absent,
which for our parameters is $\delta \theta = 0.000125$
applied every 25000 simulation time steps.
In previous work we examined lower disk densities where
the system is in the flowing state and exhibits
a series of directional locking effects
where the disks preferentially flow along specific symmetry directions of 
the obstacle lattice \cite{Reichhardt20}.
Here we focus on large obstacle sizes and/or large disk densities
where clogging effects appear.

\subsection{Directional Clogging and Memory Effect} 

\begin{figure}[h]
\centering
\includegraphics[width=3.5in]{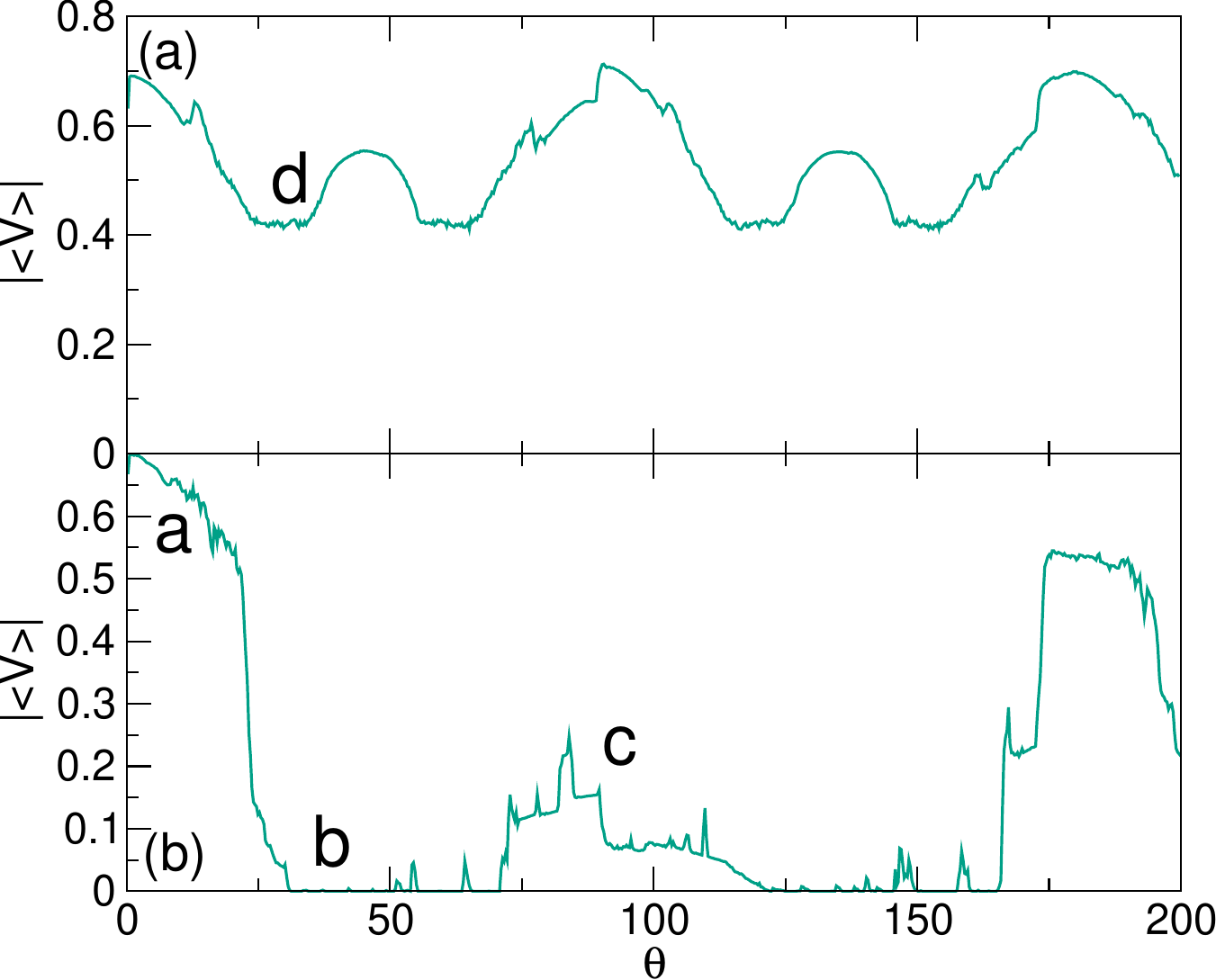}
\caption{
The instantaneous value of the total velocity $|\langle V\rangle|$
versus driving direction $\theta$
for
mobile disks moving through a square obstacle array
with $r_{\rm obs} = 1.485$
at $F_{D} = 0.0025$. 
(a) At a total system density of $\phi = 0.632$,
the disks are always flowing but $|\langle V\rangle|$ has
local maxima at $\theta = 0^\circ$, $45^\circ$ and $90^\circ$. 
(b) At $\phi = 0.68$, there are extended regions
where the system is in a clogged state.
The labels (a,b,c,d) indicate the values of $\theta$ at which
the images in Fig.~\ref{fig:2} were obtained.
}
\label{fig:1}
\end{figure}

We first consider a system with $r_{\rm obs} = 1.485$
at
$F_{D} = 0.0025$.
In Fig.~\ref{fig:1}(a) we plot $|\langle V\rangle|$ versus
$\theta$ at $\phi=0.632$ where there are $N_{\rm obs}=81$ obstacles and
$N_d=330$ mobile disks.
If the disks were flowing freely without contacting the
obstacles or other disks, we would obtain $|\langle V\rangle| = 0.825$. 
Figure~\ref{fig:1}(a) shows that $|\langle V\rangle|$ is finite for
all driving angles, indicating that the system is never in a clogged state;
however, maxima in $|\langle V\rangle|$ appear
at $\theta = 0^\circ$, $45^\circ$, and $90^\circ$.
At these symmetry directions of the substrate array,
the disks can minimize the number of collisions that occur with the
obstacles, as studied previously
\cite{Reichhardt20}.
We find a maximum value of
$|\langle V\rangle| \approx 0.7$,
corresponding to approximately
$85\%$ of the free disk velocity.   
In Fig.~\ref{fig:1}(b), we plot $|\langle V\rangle|$ versus $\theta$ for the
same system with a larger number $N_d=409$ of mobile disks, giving
$\phi = 0.68$.
There are now extended regions of
$|\langle V\rangle| = 0.0$ in which the
system is in a clogged state, such as for
$ 30^{\circ} < \theta < 70^\circ$ and $120^\circ < \theta < 167^\circ$.
Here the disks are still able to flow
along the $0^\circ$ and $90^\circ$ symmetry directions of the obstacle array
but become clogged for other driving directions.
The velocity $|\langle V\rangle| \approx 0.7$
for $\theta = 0^\circ$,
is only $|\langle V\rangle| \approx 0.15$ at $\theta=90^\circ$,
and reaches $|\langle V\rangle| \approx 0.54$ for
$\theta = 180^\circ$.
Thus the magnitude of the velocity
has a similar value for driving in the positive or negative $x$
direction but is considerably smaller for driving in the $y$ direction.
This hysteresis or memory of the initial driving direction
results from 
the fragility of the clogged states.
When the system first enters a clogged phase at
$\theta = 30^\circ$, it becomes locked to a configuration that
blocks flow for driving along or close to that particular value of $\theta$.
When $\theta$ increases to $90^\circ$, a portion of the
configuration remains
clogged 
so the flow is reduced compared to its original positive $x$ direction
value.
As $\theta$ increases to $\theta=180^\circ$, along the negative $x$
direction,
the drive exerts reversed forces 
on the configurations that formed to block the $\theta=30^\circ$ flow,
destroying these configurations and unclogging the system.
When we continue to cycle the value of $\theta$, we always find
greater flow along the $\pm x$ directions than along the $\pm y$ directions.
If we instead start the system with $\theta=90^\circ$ so that the flow is
along the $y$ direction,
we find the opposite effect in which
the flow is always higher along the $\pm y$ directions than
along the $\pm x$ directions. 
This indicates that the flow retains a memory of the initial
driving direction.
For $\phi = 0.632$ in Fig.~\ref{fig:1}(a), $|\langle V\rangle|$ exhibits
little or no memory effect since no clogging occurs, so
the values of $|\langle V\rangle|$ at $\theta = 0^\circ$ and
$\theta=90^\circ$ are nearly identical.

\begin{figure}[h]
\centering
\includegraphics[width=3.5in]{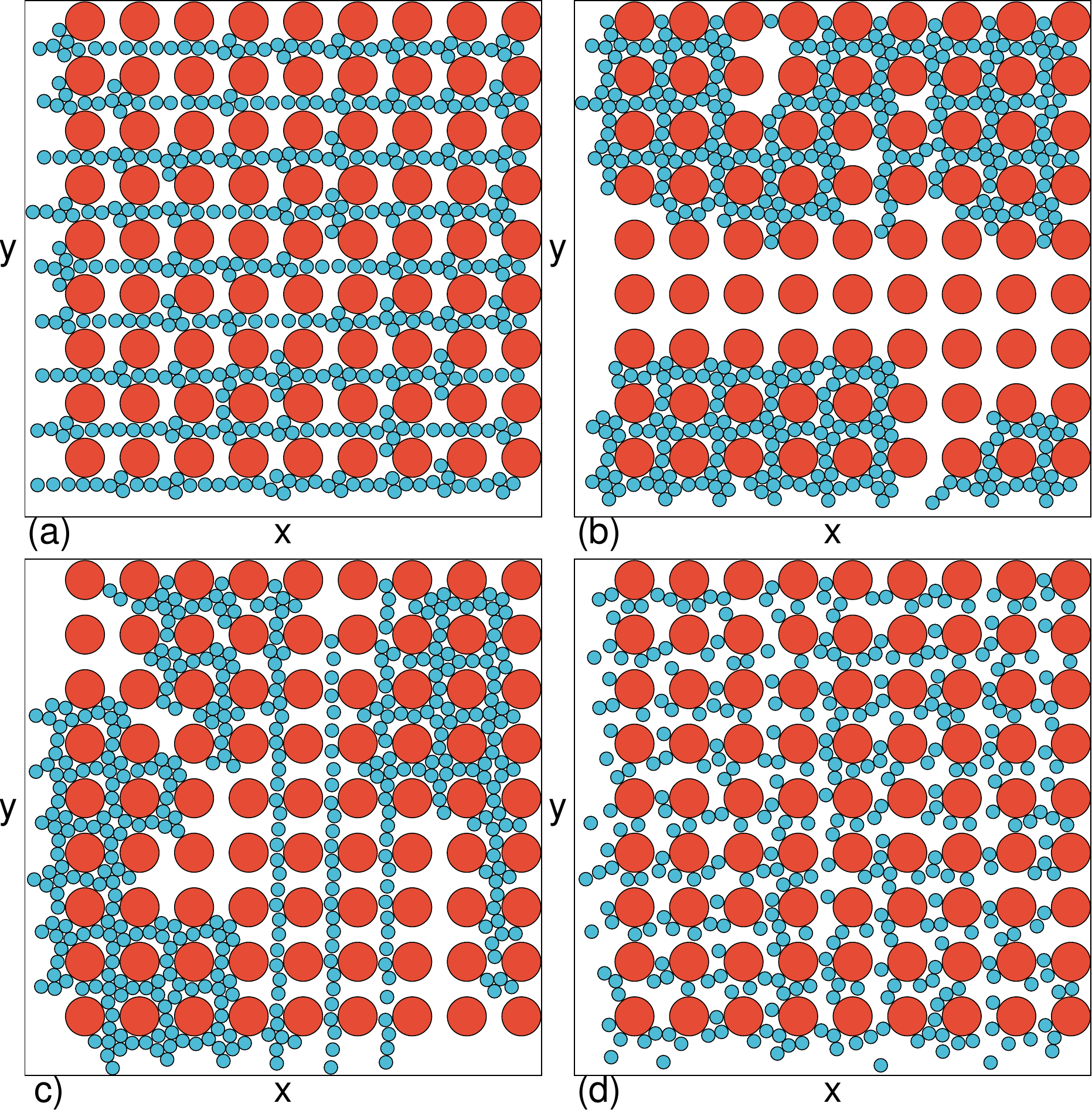}
\caption{
The obstacle locations (red circles) and mobile disks (blue circles)
for the system in Fig.~\ref{fig:1}(b) with $r_{\rm obs}=1.485$,
$F_D=0.0025$, and $\phi=0.68$ at
(a) $\theta = 2^\circ$ where the disks are flowing along the $x$
direction, (b) $\theta = 35^\circ$ where the system is in a clogged state, and
(c) $\theta = 90^\circ$ where there is a combination of clogged and
flowing disks.
(d) The same for the system in Fig.~\ref{fig:1}(a) with $\phi=0.632$
at $\theta = 35^\circ$ 
where a clogged state does not occur.
}
\label{fig:2}
\end{figure}

In Fig.~\ref{fig:2}(a) we show a snapshot of the disk
and obstacle locations for the system in
Fig.~\ref{fig:1}(b) at
$\theta = 2^\circ$ where the disks are
flowing along the $x$ direction.
Figure~\ref{fig:2}(b) illustrates
the clogged configuration at $\theta= 35^\circ$
where all the disks are immobile 
and have formed high density regions coexisting with regions that contain
no mobile disks.
In Fig.~\ref{fig:2}(c) at
$\theta = 90^{\circ}$, 
clogged configurations coexist with
moving disks which are aligned with the $y$ direction and flowing
in the driving direction
near the center of the sample.
Figure~\ref{fig:2}(d) shows the obstacle and disk configurations
at $\theta = 35^\circ$ for the system in
Fig.~\ref{fig:1}(a) where no clogged state appears and the
disk density remains uniform.

\begin{figure}[h]
\centering
\includegraphics[width=3.5in]{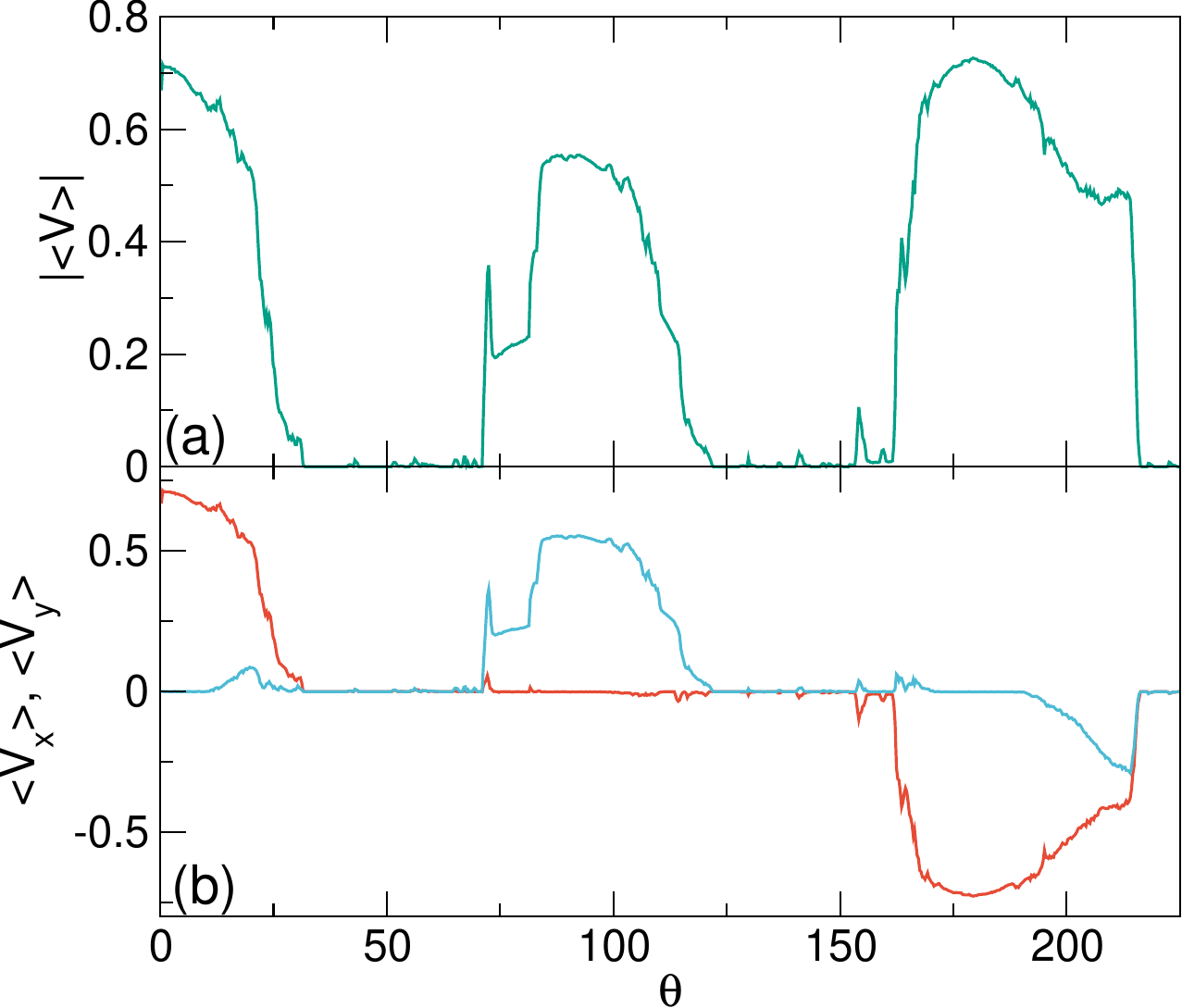}
\caption{
(a) $|\langle V\rangle|$ vs $\theta$ for the system in
Fig.~\ref{fig:1} with $r_{\rm obs}=1.485$ and $F_D=0.0025$
at a low $\phi = 0.656$ showing two clogged regimes
and a reduction of the memory,
as indicated by the fact that the velocity is nearly the same
for 
$\theta = 0^\circ$ and $\theta=90^\circ$.
(b) 
The individual velocity components
$\langle V_{x}\rangle$ (red) and $\langle V_{y}\rangle$ (blue)
vs $\theta$
showing that the flow in the non-clogged regions 
occurs preferentially along the $x$ or $y$ directions.   
}
\label{fig:3}
\end{figure}

\begin{figure}[h]
\centering
\includegraphics[width=3.5in]{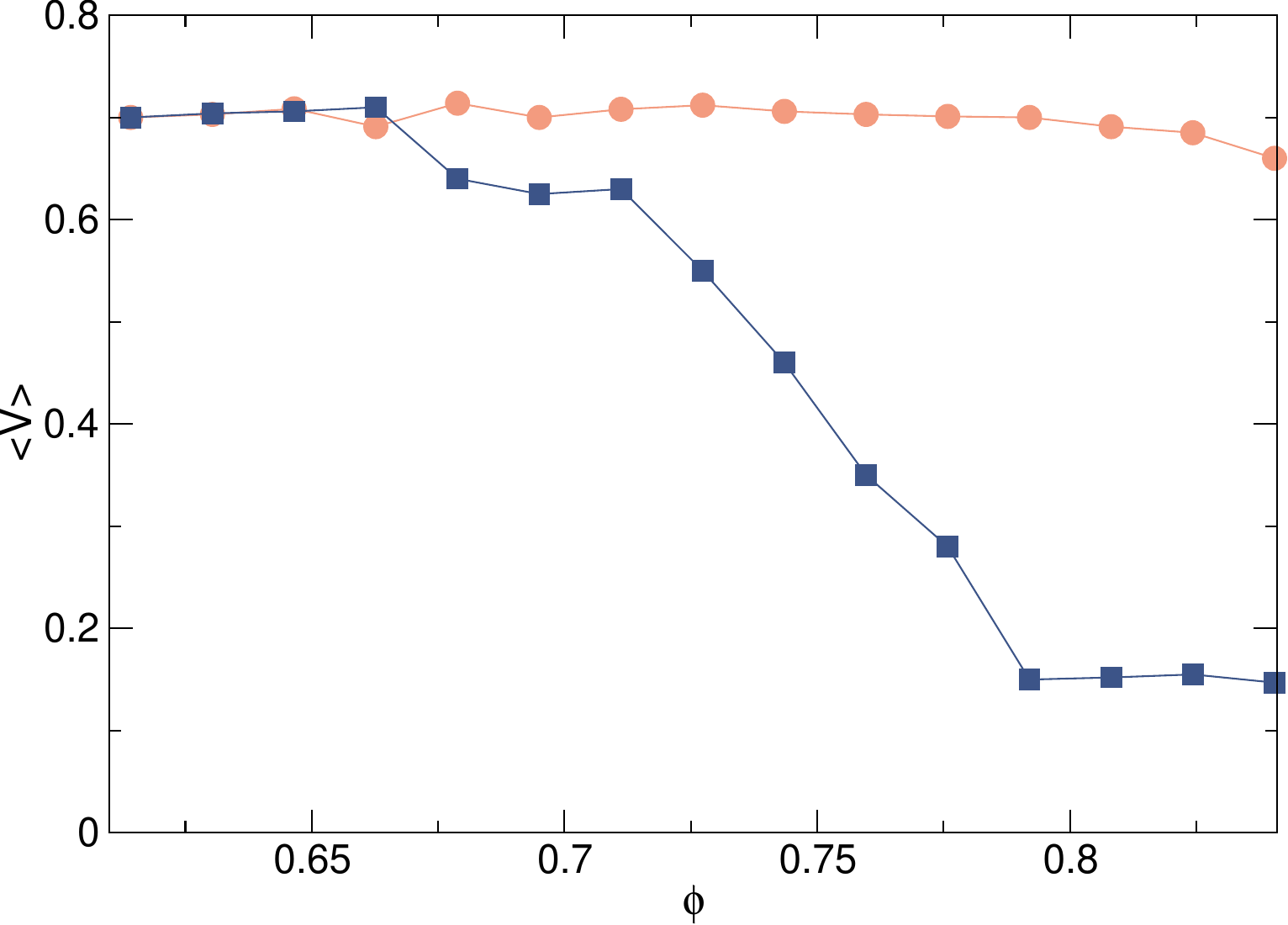}
\caption{
$\langle V\rangle$ versus density $\phi$
for the system in Fig.~\ref{fig:1} with
$r_{\rm obs}=1.485$ and $F_D=0.0025$ 
at $\theta = 0^\circ$ (orange circles)
and $\theta = 90^\circ$ (blue squares). A memory effect
in which the velocity at the two values of $\theta$ is different
appears when $\phi>0.644$.
}
\label{fig:4}
\end{figure}

As $\phi$ decreases, the clogging memory effect diminishes,
as shown in the plot of $|\langle V\rangle|$ versus $\theta$ in
Fig.~\ref{fig:3}(a)
for a sample with $\phi = 0.656$.
There are two clogged windows, but in the moving regimes,
the velocity is nearly equal in magnitude
for both $\theta = 0^\circ$ and $\theta=90^\circ$.
In Fig.~\ref{fig:3}(b) we plot the corresponding
velocity components $\langle V_{x}\rangle$ and $\langle V_{y}\rangle$
versus $\theta$.
When $\theta < 30^\circ$, the flow is predominantly
along the $x$ direction, but there is a small amount of motion
in the $y$ direction produced by the disk rearrangements 
that occur
as the system enters the clogged state.
Both velocity components are zero in the clogged regime.
For $70^\circ < \theta < 120^\circ$,
the flow is almost exclusively along the $y$ direction
since the clogged state formed for driving along the $x$ direction, 
causing motion along $x$ to be suppressed.
The memory effect continues to diminish with decreasing $\phi$
as shown in 
Fig.~\ref{fig:4} where we plot $\langle V\rangle$
versus $\phi$ at $\theta = 0^\circ$ and 
$\theta = 90^\circ$.
When the system retains a memory of the driving direction, the
net velocity for these two driving directions is different.
When
$\phi < 0.644$, clogging becomes impossible and the memory effect disappears.

\begin{figure}[h]
\centering
\includegraphics[width=3.5in]{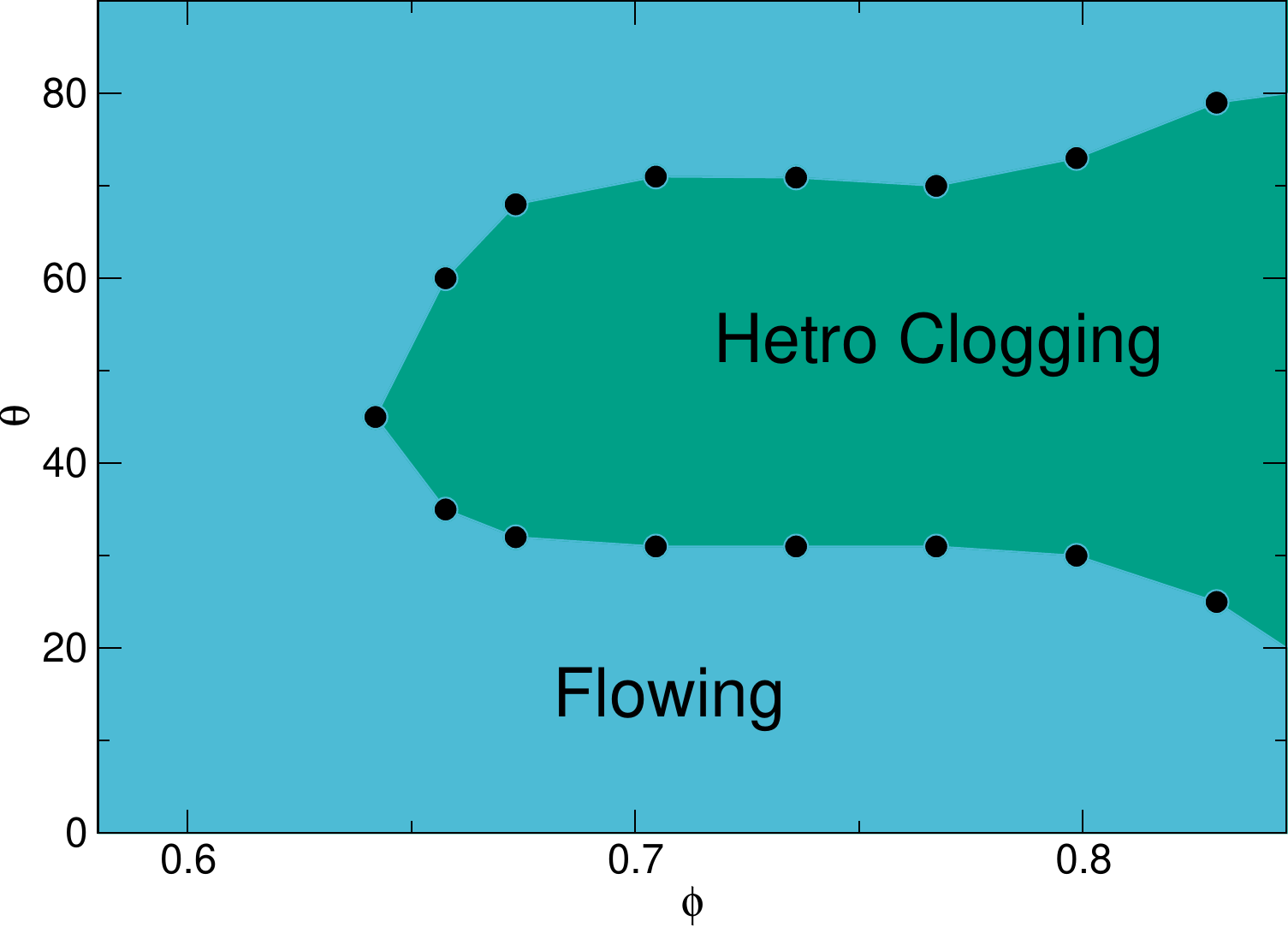}
\caption{
Dynamical phase diagram as a function of $\theta$ versus $\phi$ for the
system in Fig.~\ref{fig:1} with $r_{\rm obs}=1.485$ and $F_D=0.0025$  
showing the heterogeneous clogging regime (green) and
the flowing regime (blue).
}
\label{fig:5}
\end{figure}

The ability of the system to clog at a fixed mobile disk radius
is determined by the driving direction $\theta$,
the total density $\phi$,
and the obstacle radius $r_{\rm obs}$.
The directional dependence
arises from the changes in the effective distance
$a_{\rm eff}$ between obstacles
along the path of the mobile disks.
For $\theta=0^\circ$ and $\theta=90^\circ$,
$a_{\rm eff}$ reaches its maximum value of
$a_{\rm eff}=a$,
while at incommensurate angles, $a_{\rm eff}$ 
is reduced, permitting more frequent collisions between
the mobile disks and the obstacles.
In Fig.~\ref{fig:5} we plot a
dynamical phase diagram as a function of $\theta$ versus $\phi$
for the system in Fig.~\ref{fig:1}
highlighting where the heterogeneous clogged state appears.
For $\phi < 0.63$, the system
never clogs,
while as $\phi$ increases above $\phi=0.63$,
the width of the clogging phase increases.
For $\phi > 0.71$ our initialization
procedure cannot pack any more disks into the system; however,
we expect that for high disk densities the width
of the clogged state would continue to grow until 
the system becomes jammed for all directions of motion.
The formation of a jammed rather than a clogged state would also be
associated with the loss of the memory effect
since the velocity would be zero
for every direction of drive.
The nature of the change from clogging to jamming behavior
is beyond the scope of the present study.
The fragility that we observe in the clogged phase
is consistent with
the ideas of fragility proposed
for certain types of soft matter systems under constraints,
where special configurations or force chains
must form to block the flow in certain directions \cite{Cates98}.

\section{Clogging for Varied Obstacle Size}

\begin{figure}[h]
\centering
\includegraphics[width=3.5in]{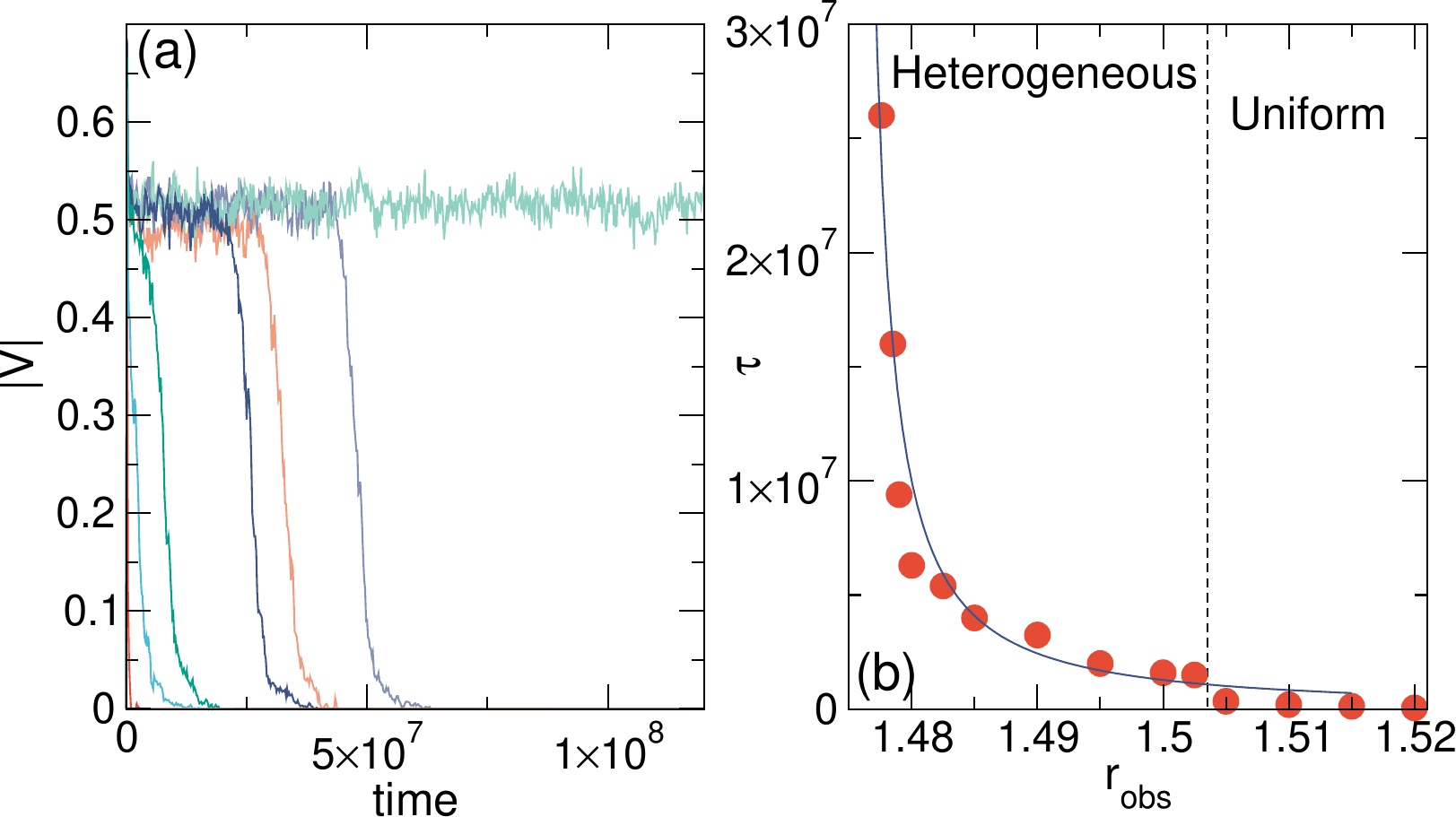}
\caption{(a) Instantaneous velocity $|V|$ versus time in simulation
time steps for the system in Fig.~\ref{fig:1} with $F_D=0.0025$ and
fixed driving angle $\theta=31^\circ$ for obstacle radius
$r_{\rm obs}=1.51$ (dark orange), $1.5025$ (light blue), $1.5$ (dark green),
$1.495$ (dark blue), $1.485$ (light orange), $1.48$ (light purple),
and $1.475$ (light green). 
(b) The time $\tau$ for the system to reach a clogged state
vs $r_{\rm obs}$. The solid line is a fit to 
$\tau \propto (r_{\rm obs} -r_{c})^{-1.25}$.
The dashed line separates the heterogeneous clogging
state from the uniform clogged state.   
}
\label{fig:6}
\end{figure}

In Fig.~\ref{fig:6}(a) we plot the
instantaneous disk velocity $|V|$ versus
time in simulation time steps for
the system in Fig.~\ref{fig:1}(b) at a fixed drive direction of
$\theta  = 31^\circ$
with varied obstacle sizes of 
$r_{\rm obs}=1.51$, $1.5025$, $1.5$, $1.495$, $1.485$, $1.48$,
and $1.475$.
The time
needed for the system to reach a zero velocity clogged state
increases
with decreasing $r_{\rm obs}$.
We find two distinct clogging regimes
as a function of $r_{\rm obs}$ for this obstacle density.
When $r_{\rm obs} > 1.502$,
the spacing between adjacent obstacles is so small that 
even a single disk
can become trapped when attempting to move between the obstacles.
The obstacle lattice constant is $a=4.0$ and the mobile disks have
radius $r_d=0.5$, so
there is only exactly enough room for the mobile disk to pass between
the obstacles without touching them
when $r_{\rm obs}=1.5$. Since the disk-obstacle interaction is represented
by a very stiff spring rather than a hard wall, disks can still slip
between the obstacles even when $r_{\rm obs}>1.5$.
For $r_{\rm obs} > 1.5025$,
the clogging occurs at the single disk level and is uniform in nature,
while for $ 1.5025 < r_{\rm obs} < 1.475$, multiple mobile disks must
interact in order to form a clogged state, resulting in spatial heterogeneity.
For $r_{\rm obs}<1.475$, the system is in a flowing state.
In the heterogeneous 
clogging regime, the system remains in its initial flowing state for
some time before
a collision between mobile disks nucleates
a high density clogged region
that can spread across the sample, blocking the flow.

\begin{figure}[h]
\centering
\includegraphics[width=3.5in]{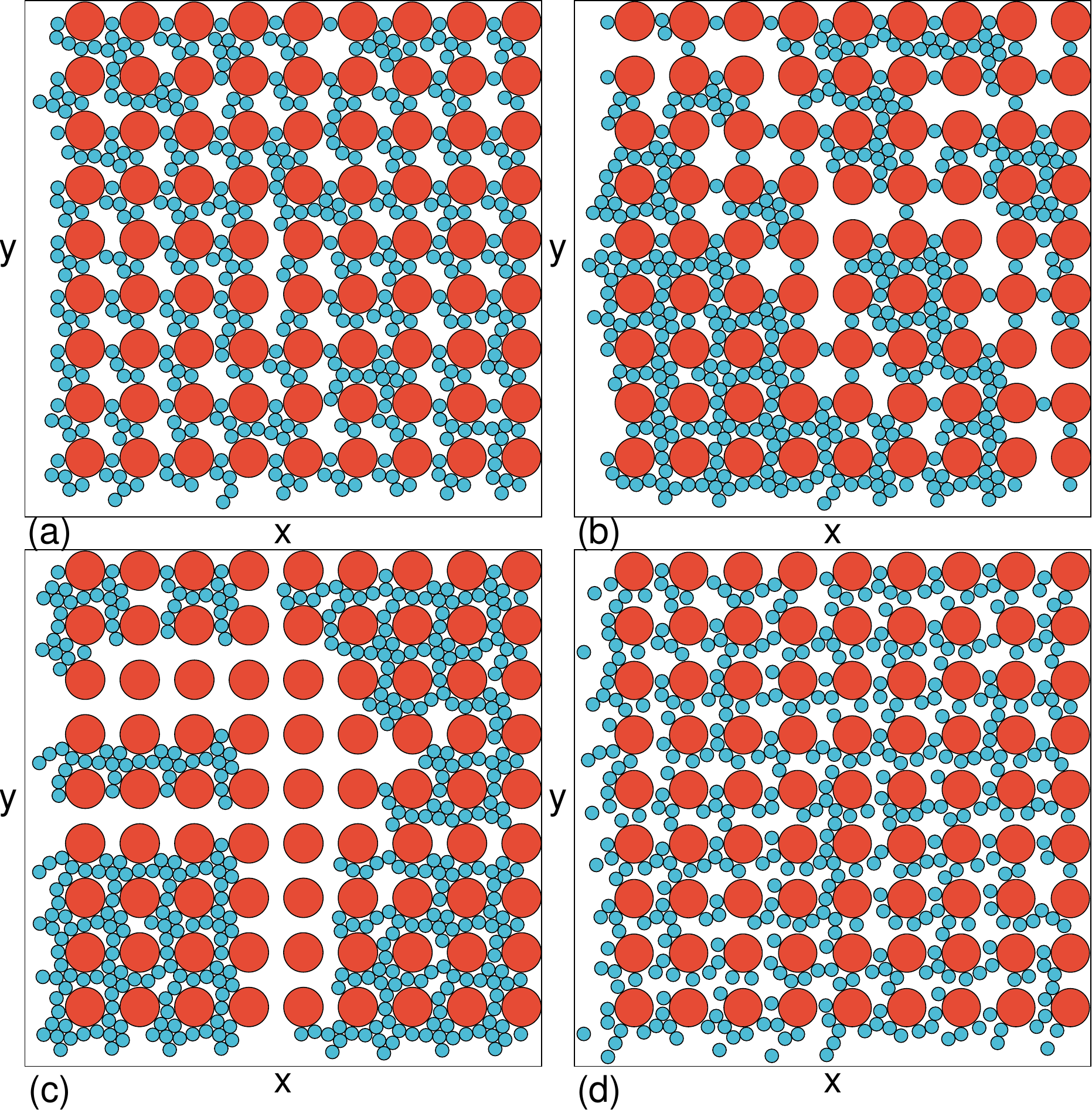}
\caption{
The obstacle locations (red circles) and mobile disks (blue circles)
for the system in Fig.~\ref{fig:6} with $F_D=0.0025$ and $\theta=31^\circ$.
(a) A uniform or homogeneous clogged state at $r_{\rm obs} = 1.51$.
(b) A clogged state for $r_{\rm obs} = 1.5025$ at the
crossover from uniform to heterogeneous clogging.
(c) A heterogeneous clogged state at $r_{\rm obs} = 1.5$.
(d) A flowing state
at $r_{\rm obs}= 1.45$.  
}
\label{fig:7}
\end{figure}

In Fig.~\ref{fig:7} we illustrate
some representative configurations for the system in
Fig.~\ref{fig:6}.
Figure~\ref{fig:7}(a) shows a uniformly clogged state 
at $r_{\rm obs} = 1.51$, where the disks
are all immobile but the density is uniform.
At $r_{\rm obs}=1.5025$ in
Fig.~\ref{fig:7}(b), there is a transition to heterogeneous clogging.
In Fig.~\ref{fig:6}(c) we show a heterogeneous clogged state at
$r_{\rm obs} = 1.5$,
while at $r_{\rm obs}=1.45$ in Fig.~\ref{fig:6}(d),
the disks are flowing.

We measure the time $\tau$ required for the system to reach a clogged state by
fitting the curves in Fig.~\ref{fig:6}(a) to the form
$|V|(t) \propto \exp{-t/\tau} + V_0$.
In Fig.~\ref{fig:6}(b) we plot  
$\tau$ versus $r_{\rm obs}$,
showing a divergence
near a critical obstacle radius of $r_{c}  = 1.4752$.
The solid line is a fit to 
$\tau \propto (r_{\rm obs} -r_{c})^{\gamma}$ with $\gamma=-1.25$.
When $r_{\rm obs} > 1.5025$, $\tau$ drops to a small value
since no plastic rearrangements are required for the system to reach
a uniform clogged state. The dashed line indicates the transition from the
heterogeneous to the uniform clogging behavior.
The power law divergence in $\tau$ near $r_c$ resembles
the time divergence found at reversible to irreversible
transitions
in periodically sheared colloidal systems \cite{Corte08,Milz13},
amorphous solids \cite{Regev13}, and
superconducting vortices \cite{Okuma11}.
In previous numeral work on clogging in two-dimensional
random obstacle arrays \cite{Peter18},
a similar power law time divergence
with an exponent of $\gamma=-1.29$ appeared when
the system entered the clogged phase as the obstacle density was varied.
These exponents 
are close to the
value expected for an absorbing phase transition,
where the clogged state can be viewed as the absorbed 
state since in this state all fluctuations are lost \cite{Hinrichsen00}.

When the obstacles are in a periodic array, the nature of the clogged
state
depends on the driving direction.
For $\theta = 0^\circ$ 
or $\theta = 90^\circ$, there is only a uniform clogged state
for
$r_{\rm obs} > 1.5025$ but there are no heterogeneous clogged states,
so we find no power law divergence in the clogging time
for these driving directions.
At incommensurate angles,
the system has a closer resemblance to a random obstacle array,
making it possible for
a heterogeneous clogged state to appear that is associated with
a power law divergence in the time required for the clogged state to
organize.
In this work we focus only on monodisperse mobile 
disks,
but if the mobile disk radii were bidisperse or multidisperse,
the system could exhibit heterogeneous clogging 
for $x$ and $y$-direction driving.
In this case, the clogging transition would likely shift to
lower values of $\phi$ and $r_{\rm obs}$.

\begin{figure}[h]
\centering
\includegraphics[width=3.5in]{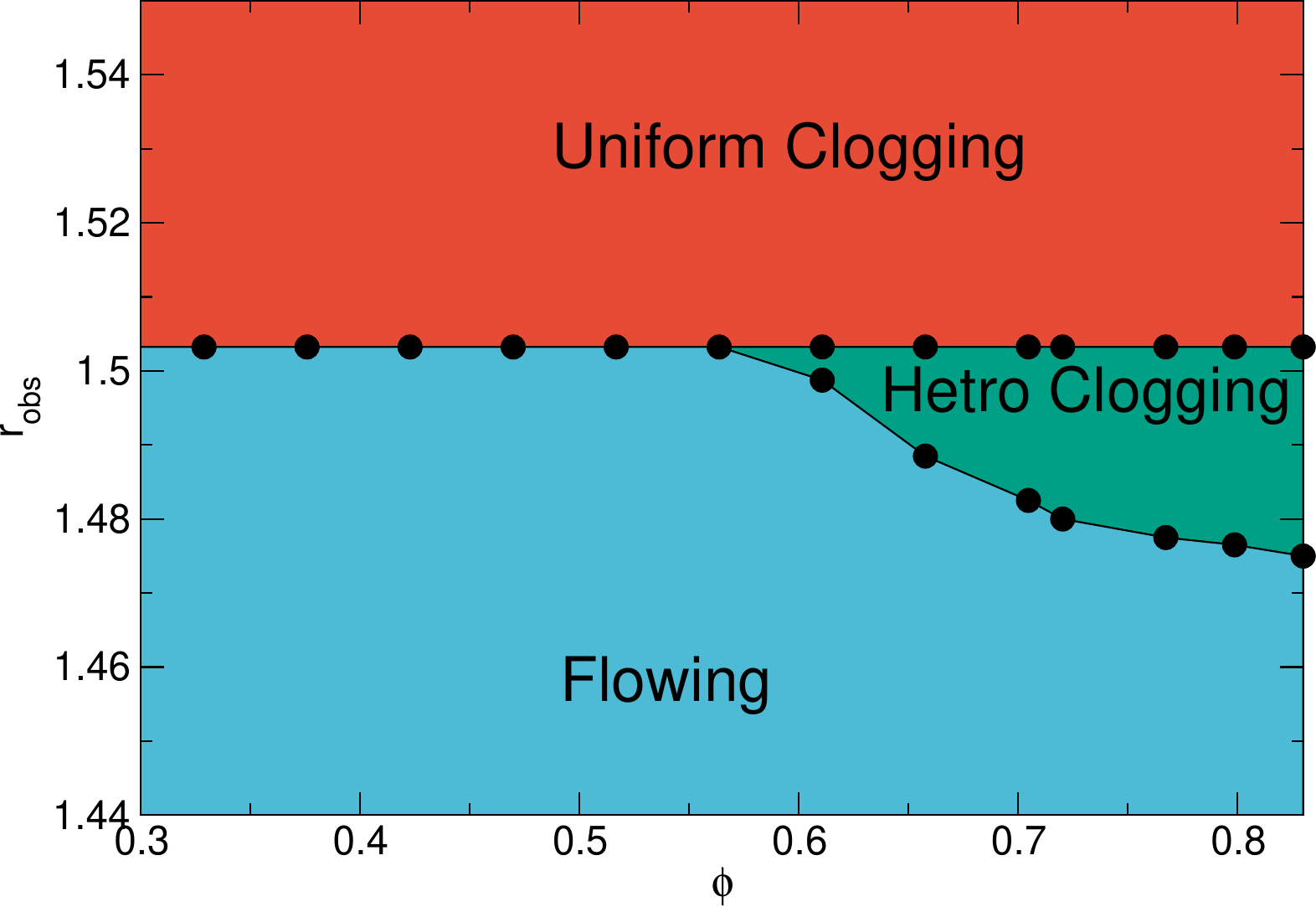}
\caption{
Dynamical phase diagram as a function of $r_{\rm obs}$ vs $\phi$ for the system
in Fig.~\ref{fig:6} with $F_D=0.0025$ and fixed driving angle $\theta=31^\circ$
showing the heterogeneous clogging regime (green), flowing regime (blue),
and uniform clogging regime (red).
}
\label{fig:8}
\end{figure}

In Fig.~\ref{fig:8} we plot a dynamical phase diagram as a function
of $r_{\rm obs}$ versus $\phi$ for the system in Figs.~\ref{fig:6} and
\ref{fig:7} where the drive is applied at $\theta=31^\circ$.
For $r_{\rm obs} > 1.503$, the system forms a uniform clogged state that
is independent of $\phi$, and the clogged state forms immediately with
no diverging time scale.
When $\phi > 0.6$ and $1.475 < r_{\rm obs} < 1.5032$,
we find heterogeneous clogging with a power law time divergence
for the formation of the clogged state.
Similar phase diagrams can be constructed for other
driving angles. For example, at $\theta = 0^\circ$ the heterogeneous
clogged phase is absent for the same range of system parameters
shown in Fig.~\ref{fig:8}.

In our studies we have not considered
the effect of temperature or other perturbations such as activity
\cite{Reichhardt18a,Ai19}. Such perturbations are
likely to wash out the clogged state due to its fragile nature;
however, there could still 
be some remnant of nonlinear behavior or intermittent dynamics in
regions where heterogeneous clogging would occur in the absence of the
perturbations.  

\section{Depinning of the Clogged Phase}
Since the disk-disk interactions in our system have a harmonic form,
the clogged phase should exhibit 
a drive dependence or a critical driving
force above which it should unclog or depin.
This type of depinning or unclogging effect is applicable to
systems such as bubbles, emulsions, soft colloids, or magnetic bubbles.
On the other hand, in granular matter or other systems with hard core
particle-particle interactions where the particles cannot deform easily,
such depinning would likely occur only for much stronger drives and
would be difficult to access experimentally.

\begin{figure}[h]
\centering
\includegraphics[width=3.5in]{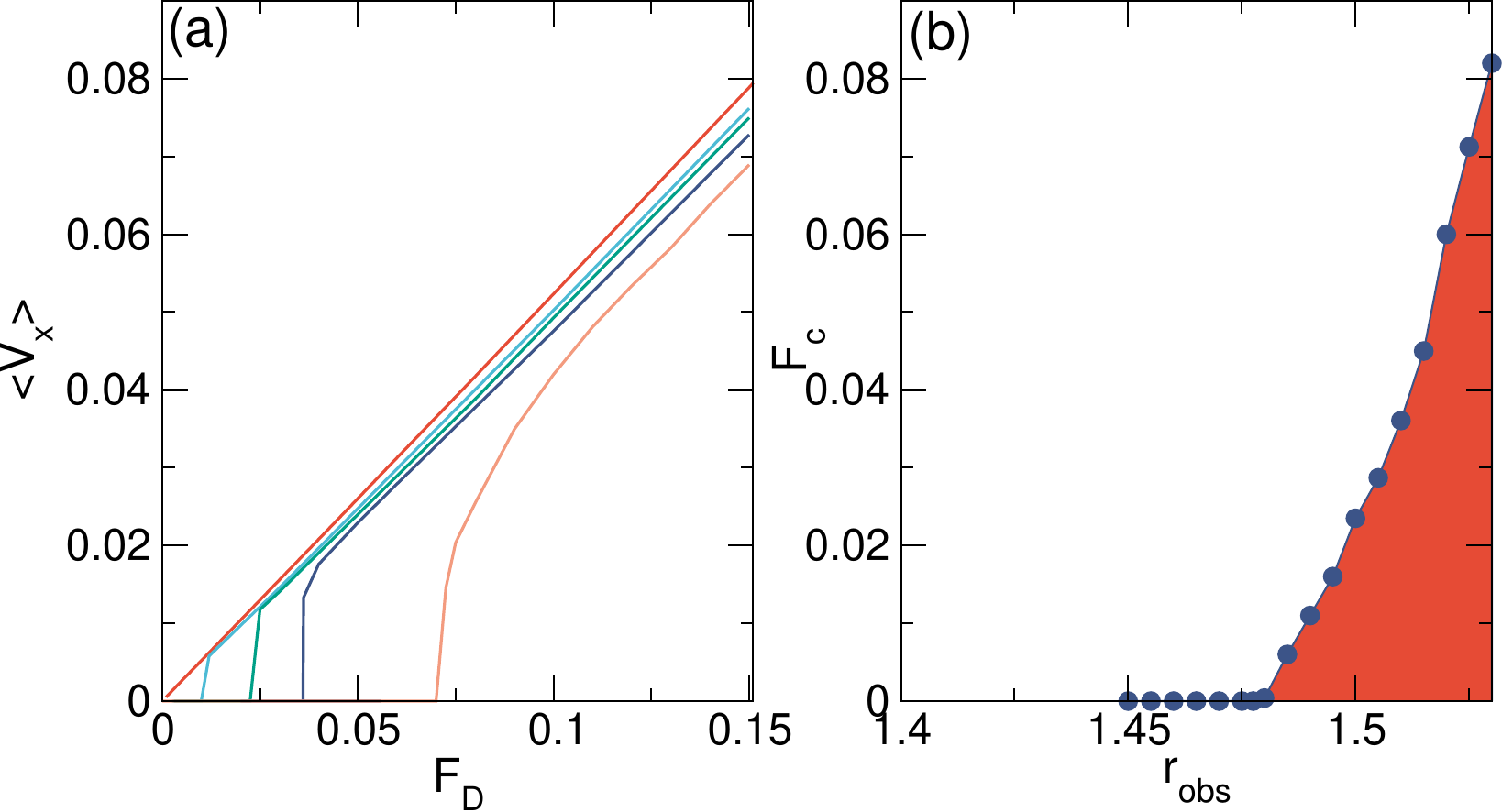}
\caption{
(a) The average velocity $\langle V\rangle$ vs
$F_{D}$ for the system in Fig.~\ref{fig:1}
with $\phi=0.68$ and
a drive angle of $\theta = 33.8^\circ$ for
$r_{\rm obs}= 1.525$ (orange), $1.51$ (dark blue),
$1.5$ (green), $1.49$ (light blue), and $1.475$ (orange red).
(b) The depinning threshold $F_{C}$
vs $r_{\rm obs}$ for the system in (a). 
}
\label{fig:9}
\end{figure}     

\begin{figure}[h]
\centering
\includegraphics[width=3.5in]{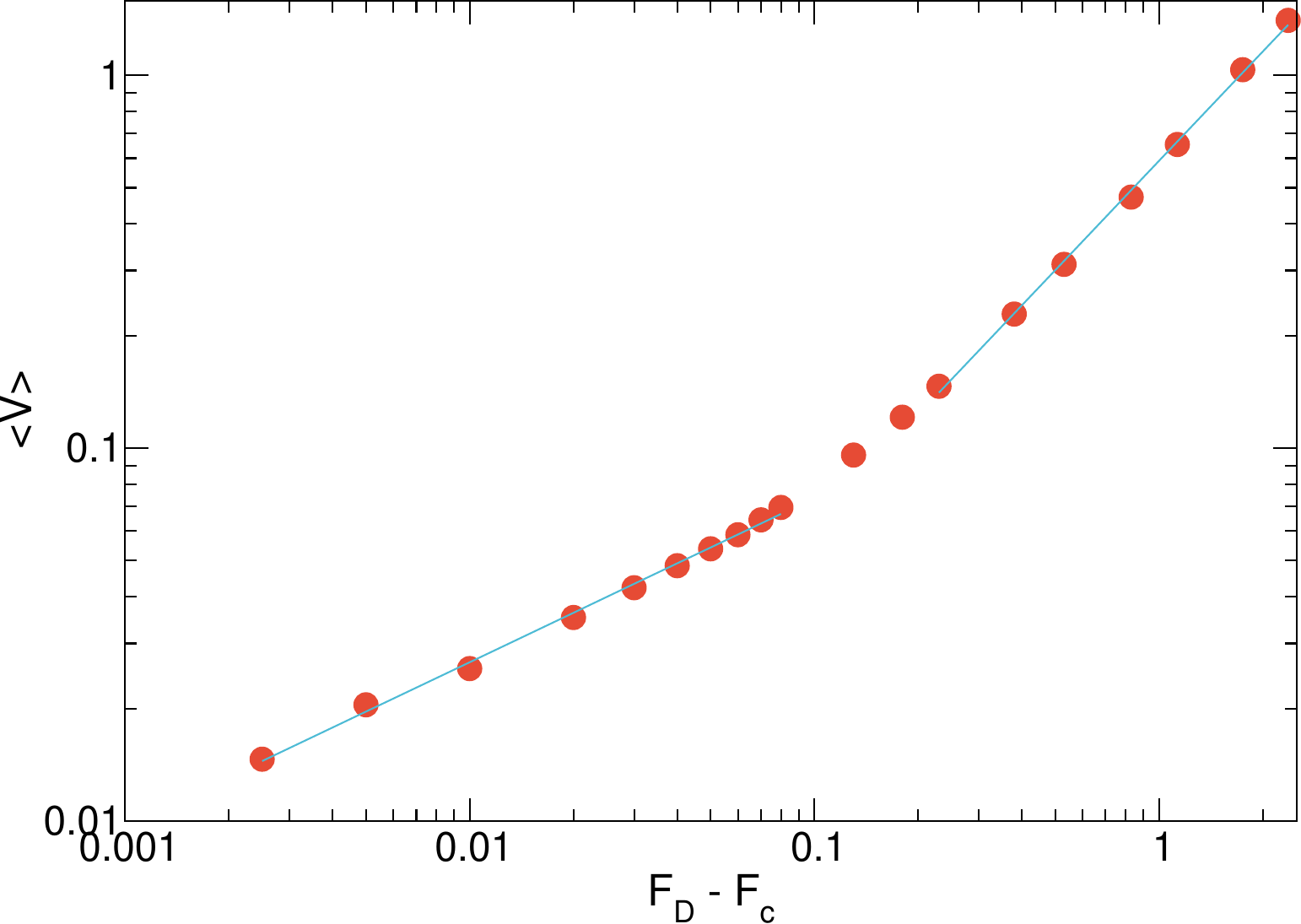}
\caption{ 
The scaling of the velocity-force curve from
Fig.~\ref{fig:9}(a) plotted as $\langle V\rangle$ vs $F_D-F_c$
in the uniform clogged phase at $r_{\rm obs} = 1.525$, where $F_C=0.07$.
Here $\phi=0.68$ and $\theta=33.8^\circ$.
The leftmost solid line is a power law fit with
$\beta = 0.44$, while at higher
drives, there is a crossover to a linear behavior with $\beta = 0.97$,
as indicated by the rightmost solid line.
}
\label{fig:10}
\end{figure} 

In the previous sections, we considered a drive force of
$F_{D} = 0.0025$ which is well below the depinning threshold.
We now sweep the value of $F_D$ to explore the depinning behavior.
In Fig.~\ref{fig:9} we plot $\langle V\rangle$
versus $F_{D}$ for the system in Fig.~\ref{fig:1}
at $\phi=0.68$ and
a drive angle of $\theta = 33.8^\circ$
for $r_{\rm obs}= 1.525$, $1.51$, $1.5$, $1.49$, and $1.475$.
When $r_{\rm obs}>1.475$,
there is a finite depinning threshold $F_{c}$ 
which increases with increasing $r_{\rm obs}$.
At $r_{\rm obs} = 1.525$, the velocity-force curve has
an upward concavity
and can be fit to the form
$V = (F_{D} - F_{c})^\beta$ with $\beta = 0.44$,
as shown in the left side of Fig.~\ref{fig:10}. 
In general, systems that exhibit elastic depinning
have a depinning exponent of $\beta < 1.0$ \cite{Reichhardt17}.  
During the depinning process from the clogged state,
the disks maintain the same neighbors and there is no plastic flow.
At higher drives, the velocity crosses over to a linear form with
$V \propto F_{D}$,
as shown in the right side of Fig.~\ref{fig:10}
which illustrates a fit with $\beta = 0.97$.
We generally find that depinning in the uniform
clogging phase is elastic, and that the disk density remains
uniform in both the pinned and flowing states.
Depinning in the
heterogeneous clogged phase is more consistent with
a discontinuous jump,
which could be indicative of a first order type of transition.
Here the pinned state is phase separated but the flowing state
has a uniform disk density.
This result is consistent 
with work on the depinning of two-dimensional phase separated systems
which has a first order character, both when
the pinned state is phase separated and
the flowing state is uniform
or when the pinned state is uniform and
the flowing state is phase separated \cite{Reichhardt03a,Reichhardt03}. 
In Fig.~\ref{fig:9}(b) we plot the depinning threshold
$F_{c}$ versus $r_{\rm obs}$ for the system in Fig.~\ref{fig:9}(a),
showing that there 
is no depinning threshold when $r_{\rm obs} < 1.475$. 

\subsection{Fluctuations} 

\begin{figure}[h]
\centering
\includegraphics[width=3.5in]{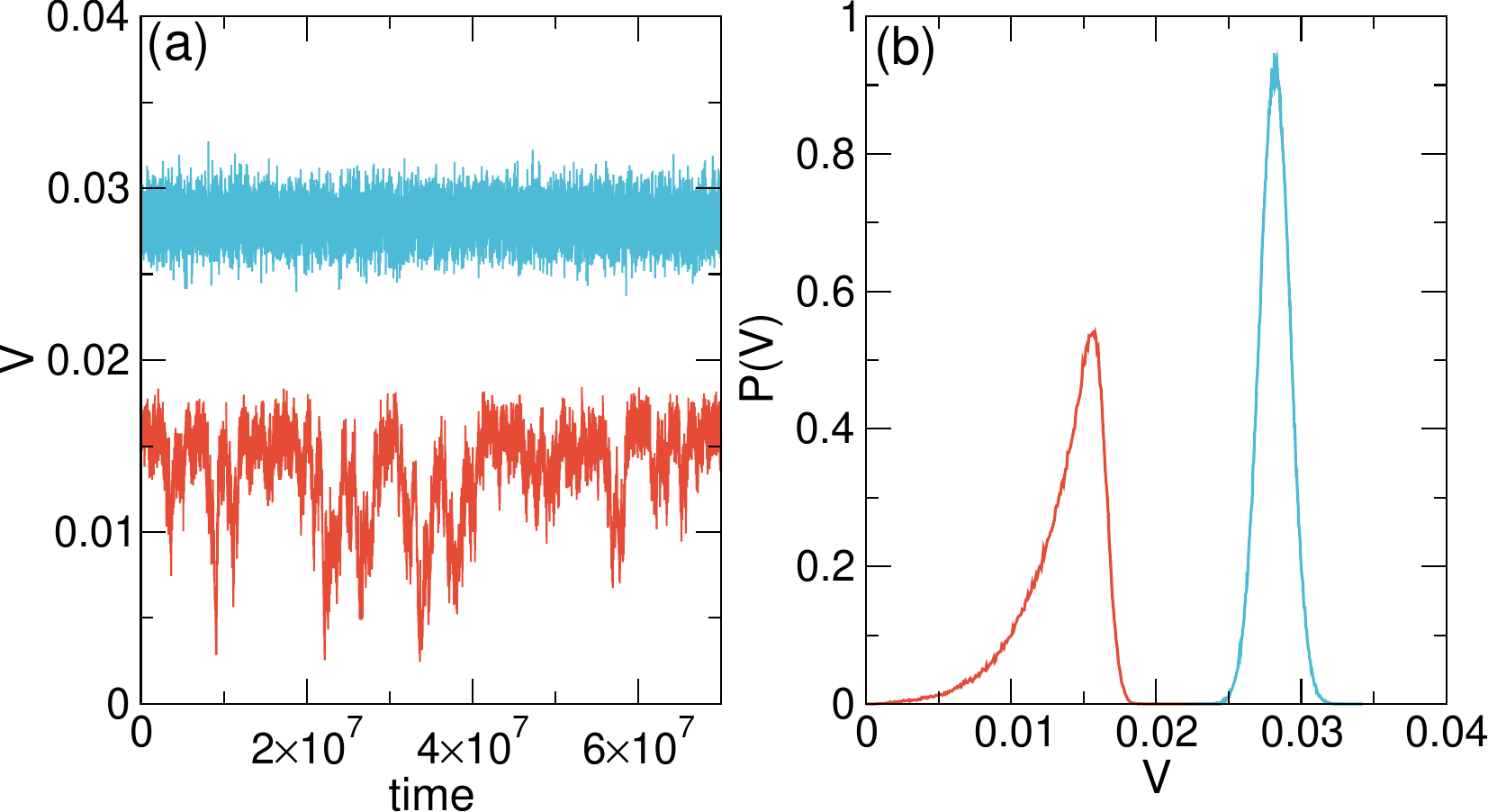}
\caption{ 
(a) Instantaneous velocity $V$ vs time in simulation time steps
for the system in Fig.~\ref{fig:9}(a)
with $\phi=0.68$, $\theta=33.8^\circ$, and
$r_{\rm obs} = 1.51$ at
$F_{D}/F_{C} = 1.0034$ (red) 
and $F_{D}/F_{C} = 1.67$ (blue).
(b) The corresponding velocity distributions $P(V)$.
}
\label{fig:11}
\end{figure}

We next address the nature of the fluctuations of the flow above
the declogging force $F_{c}$.
Generally we observe highly intermittent flow
immediately above the declogging or depinning transition,
where regions which are temporarily clogged coexist
with moving or flowing regions, while
at higher drives all of the disks are flowing.
In Fig.~\ref{fig:11}(a) we
plot the instantaneous velocity $V$ versus time for 
the system in Fig.~\ref{fig:9}(a) with
$r_{\rm obs} = 1.51$ in the uniform clogged phase for
$F_{D}/F_{C} = 1.0034$, just above the depinning threshold, as well
as for a higher drive of $F_{D}/F_{C} = 1.67$.
There are strong fluctuations in $V$
just above the depinning threshold,
while at the higher drive the velocity variations are reduced.
The fluctuations near the depinning threshold
are strongly non-Gaussian,
as shown in Fig.~\ref{fig:11}(b) where we plot $P(V)$ for the samples
in Fig.~\ref{fig:11}(a).
For $F_{D}/F_{C} = 1.0034$, $P(V)$ has
an enhanced tail 
at lower drives, producing a strongly skewed distribution,
while for $F_{D}/F_{C} = 1.67$, $P(V)$ has a more symmetrical Gaussian 
shape. We observe similar trends for the other values of $r_{\rm obs}$.

\begin{figure}[h]
\centering
\includegraphics[width=3.5in]{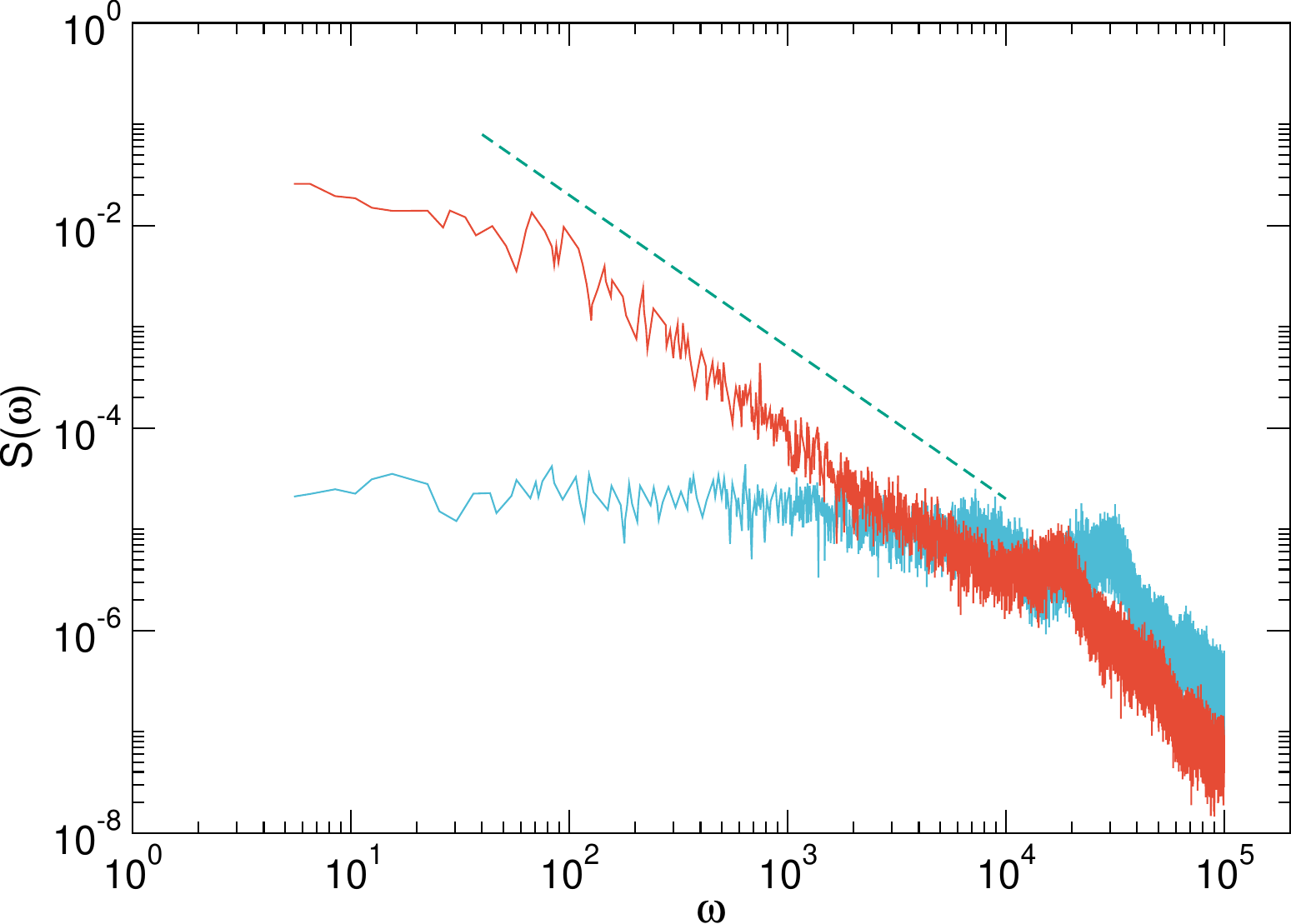}
\caption{ 
The power spectra $S(\omega)$ vs $\omega$
for the system in Fig.~\ref{fig:11}(a) with
$\phi=0.68$, $\theta=33.8^\circ$, and $r_{\rm obs}=1.51$ at
$F_{D}/F_{C} = 1.0034$ (red) and $F_D/F_C=1.67$.
The dashed line is a power law fit of the $F_D/F_C=1.0034$ curve 
to $\alpha = -1.5$, while
the $F_{D}/F_{C} = 1.67$ curve
exhibits white noise at lower frequencies with $\alpha = 0$.  
}
\label{fig:12}
\end{figure}

Differences in the noise fluctuations can also be detected by
computing
the power spectrum of the velocity time series,
$S(\omega) = |\int \exp(-2\pi i \omega) V(t)dt|^2$.
In Fig.~\ref{fig:12} we plot $S(\omega)$ for the system in Fig.~\ref{fig:11}(a).
For 
$F_{D}/F_{C} = 1.0034$,
the fluctuations have a $1/f^\alpha$ or broad band noise character with 
$\alpha = -1.5$,
while for $F_{D}/F_{C} = 1.67$,
we find a white noise signature with $\alpha = 0$.
There are peaks in $S(\omega)$ at higher $\omega$
produced by the periodic signal from the disks encountering the obstacle
lattice.
The lower frequency $1/f^\alpha$ noise is associated
with long time large scale changes
in the disk configurations.
Even under very strong fluctuations, the velocity above the depinning
transition never drops to zero because this would cause the system to be
permanently captured in a clogged state.
In contrast, other systems
with constant flux or some periodic perturbation
would show
intermittent flow
that would be expected to have $1/f^\alpha$ noise characteristics.
Studies of clogging in bottlenecks have also found
strongly intermittent dynamics
including power law distributions of bursts \cite{Gella19,Souzy20}. 

\section{Clogging in Diluted Arrays}

\begin{figure}[h]
\centering
\includegraphics[width=3.5in]{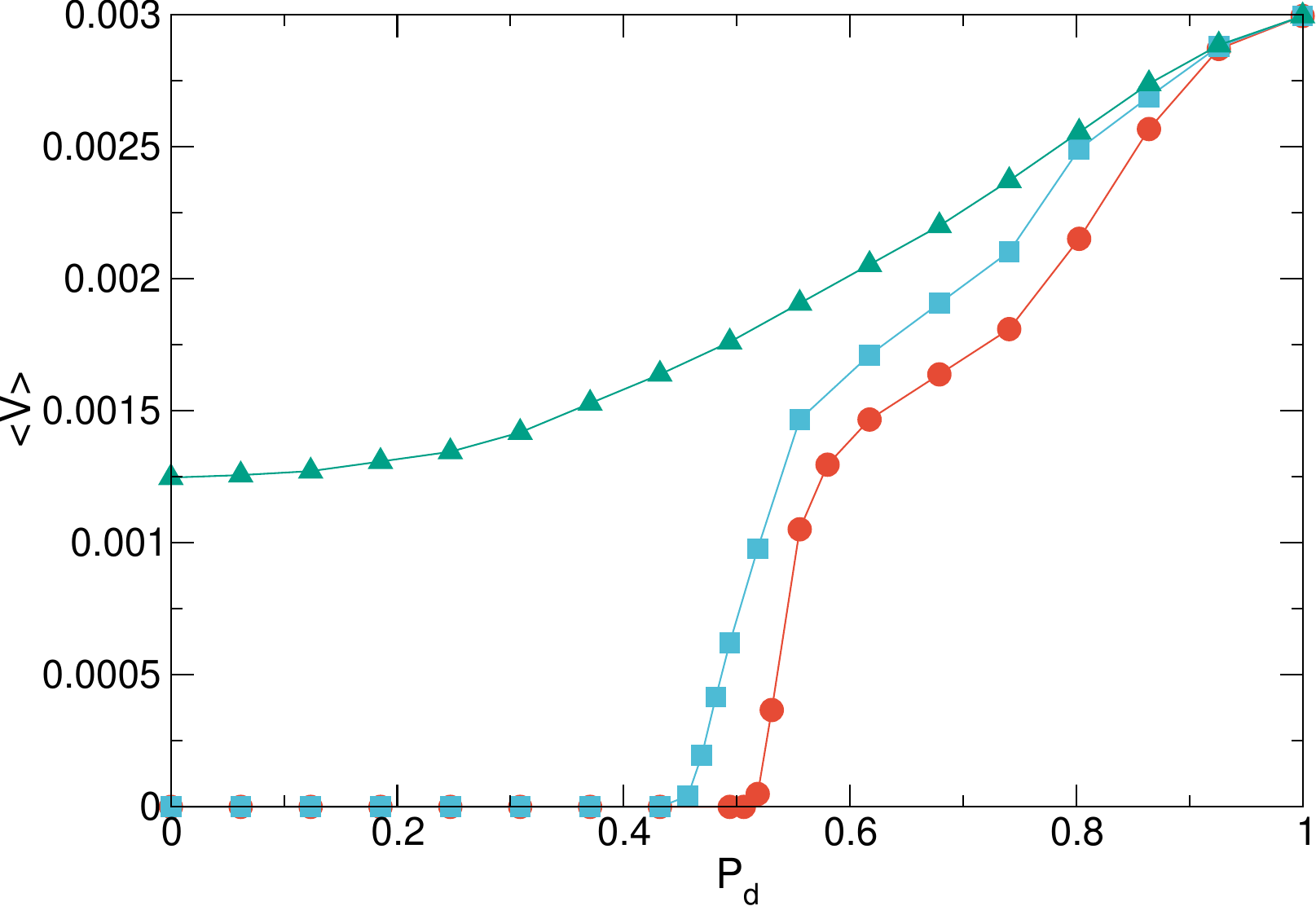}
\caption{ 
The net velocity $\langle V\rangle$ vs the dilution factor $P_d$
for random dilution of the square array
in the system from Fig.~\ref{fig:9} with
an undiluted value of $\phi=0.68$, $\theta = 33.8^\circ$ and $F_{D} = 0.0029$
at $r_{\rm obs} = 1.525$ (red circles), $1.485$ (blue squares), and 
1.475 (green triangles).  
}
\label{fig:13}
\end{figure}

In the absence of obstacles, the disks would flow for any finite drive.
Thus we study random dilution of the obstacle array in order to observe
the transition from a clogged state to a flowing phase.
We select a fraction $P_d$ of obstacles to remove at random from the system
in Fig.~\ref{fig:9} with
an undiluted value of $\phi=0.68$, $\theta=33.8^\circ$, and
$F_D=0.0029$, well below the depinning threshold
of the undiluted sample.
In Fig.~\ref{fig:13} we plot $\langle V\rangle$
versus the dilution fraction $P_d$ for
$r_{\rm obs} = 1.525$, $r_{\rm obs} = 1.485$, and $r_{\rm obs} = 1.475$.
The $r_{\rm obs} = 1.475$ has no depinning threshold even when
$P_d = 0$, and as $P_d$ increases,
there is a gradual increase in $\langle V\rangle$
which reaches a saturation value of $F_D$ near $F_d = 1.0$. 
In the $r_{\rm obs} = 1.525$ sample,
the depinning threshold 
for $P_d=0$ is $F_{C} = 0.07$.
As $P_d$ increases, the system remains clogged
up to $P_d = 0.52$, and then there is a gradual
increase in $\langle V\rangle$ as the dilution fraction becomes larger.
In general, we find that when the $P_d=0$ depinning
threshold is finite,
the dilution needs to be greater than $P_d = 0.44$ in order to unclog the
system, as shown for the $r_{\rm obs} = 1.485$ sample.
This indicates that the transition from a clogged to a flowing state
is probably related to a percolation transition.

\begin{figure}[h]
\centering
\includegraphics[width=3.5in]{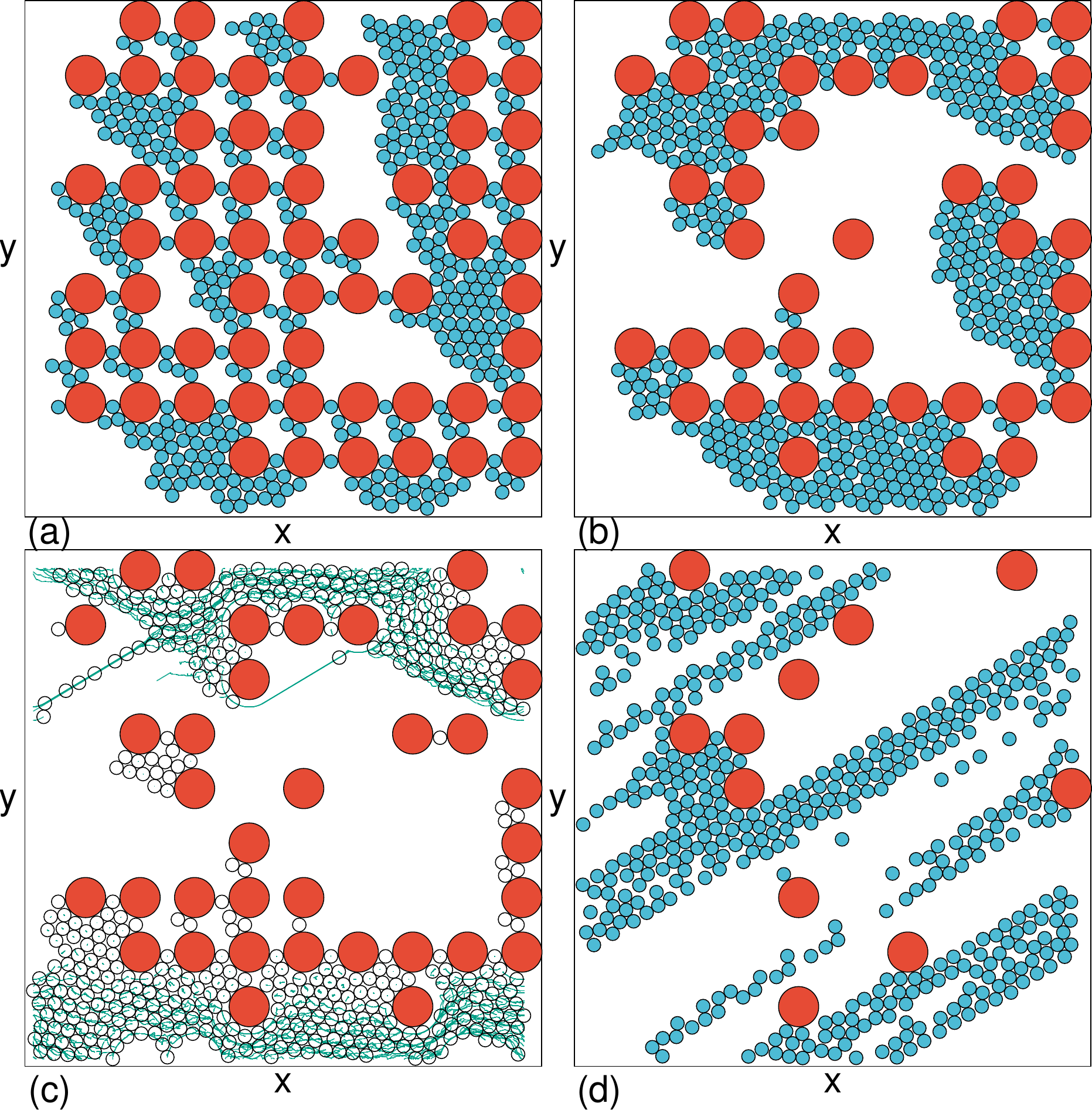}
\caption{
The obstacle locations (red circles) and mobile disks (blue circles) 
for the system in Fig.~\ref{fig:13}
with an undiluted value of $\phi=0.68$, $F_D=0.0025$ and $\theta=31^\circ$
at $r_{\rm obs} = 1.525$ under different pinning dilutions $P_d$.
(a) The clogged phase at 
$P_d = 0.25$. (b) The clogged phase at $P_d = 0.494$.
(c) The moving phase at $P_d = 0.56$ where green lines indicate the
trajectories of the mobile disks.
(d) The heterogeneous moving phase at $P_d = 0.86$.
}
\label{fig:14}
\end{figure}

\begin{figure}[h]
\centering
\includegraphics[width=3.5in]{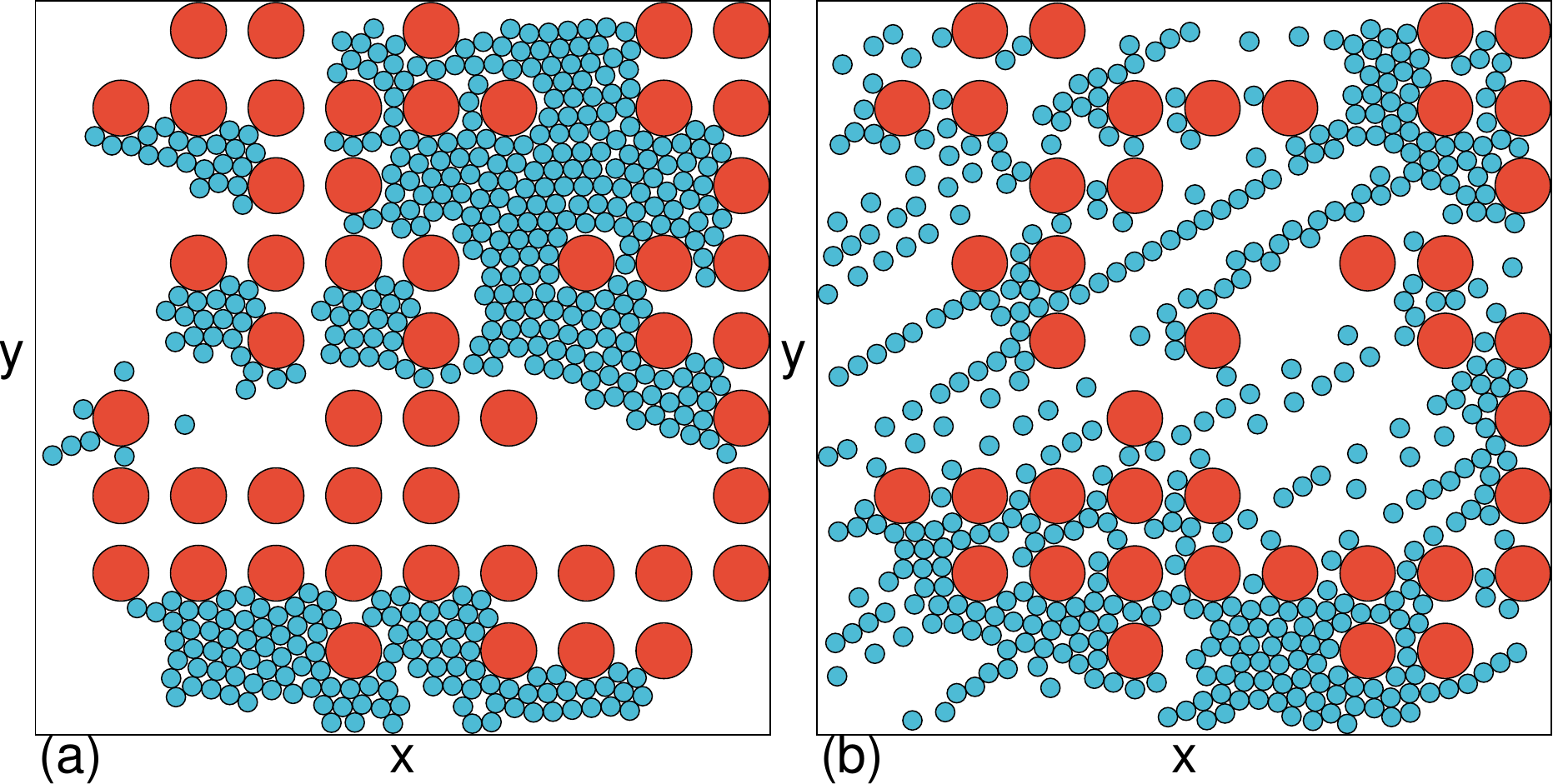}
\caption{ 
The obstacle locations (red circles) and mobile disks
(blue circles) for the system in
Fig.~\ref{fig:13} with an undiluted value of $\phi=0.68$,
$F_D=0.0025$, and $\theta=31^\circ$.
(a) At $r_{\rm obs} = 1.485$ and 
$P_d = 0.37$ in a barely clogged state, the disk arrangement
is strongly heterogeneous.
(b) A flowing state at $r_{\rm obs} = 1.475$ and $P_d = 0.494$.  
}
\label{fig:15}
\end{figure}

As the dilution is
increased, the time required for the system to organize to a
steady state increases but shows strong fluctuations if
different initializations of the mobile disk locations are used.
When $r_{\rm obs} = 1.525$, the $P_d = 0$ sample forms
a uniform clogged state; however, as the dilution increases
up to $P_d=0.5$, the clogged state becomes increasingly heterogeneous.
This is illustrated in
Fig.~\ref{fig:14}.
At $P_d = 0.25$ in Fig.~\ref{fig:14}(a),
the system is strongly spatially heterogeneous but is still clogged.
The clogged state that appears at $P_d=0.494$ just before the transition
to a moving phase is shown in Fig.~\ref{fig:14}(b).
For  $0.52 < P_d < 0.62$, clogged regions coexist
with moving regions, resulting in plastic flow as indicated
in 
Fig.~\ref{fig:14}(c) at $P_d = 0.56$.
At high dilution, the system forms a moving phase that is
distinguished from the
moving states found in undiluted arrays by its strong spatial
heterogeneity,
as shown in Fig.~\ref{fig:14}(d) for a sample with $P_d = 0.86$. 
We observe similar dynamics in the diluted systems whenever the depinning
threshold is finite.
In Fig.~\ref{fig:15}(a) we plot a barely clogged
configuration at $r_{\rm obs} = 1.485$ and a dilution of $P_d=0.37$, 
showing a heterogeneous spanning clogged state, while
in Fig.~\ref{fig:15}(b) we illustrate the moving state at
$r_{\rm obs} = 1.475$ and $P_d = 0.49$,
where a few regions are locally clogged
but the system remains in a flowing state.

\section{Conclusions}
We have examined the clogging dynamics for a monodisperse assembly of disks
moving through a periodic 
obstacle array. We find that the susceptibility for the system to
clog under fixed disk density and obstacle radius
depends on the direction of drive
relative to the symmetry of the obstacle lattice.
The
system clogs at incommensurate driving angles or
for angles in the range $30^\circ < \theta < 70^\circ$;
however, the range of parameters over which clogging occurs
increases with increasing system density and obstacle size.
The systems is least susceptible to clogging for 
drives centered around $\theta=0^{\circ}$ and $\theta=90^\circ$. 
Under a changing drive angle the
system exhibits a memory effect in which the formation of a clogged state for
one driving direction results in a reduced flow rate when the drive
is rotated into the perpendicular direction.
The memory effect is lost as the disk density or obstacle radius
decreases.
We observe two distinct types of clogging states:
heterogeneous or phase separated clogging
in which groups of disks must gradually arrange themselves into a
clogged configuration, and
a uniform clogged state in which the spacing between adjacent obstacles is
small enough that individual disks can be trapped immediately.
Since we represent the disk-disk interactions with a stiff harmonic potential,
a clogged state
can be unclogged by increasing the driving force and inducing a
depinning transition.
The disk configurations are generally uniform in the unpinned phase.
For drives just above the unclogging transition, the
velocity exhibits strong non-Gaussian fluctuations
with a $1/f^\alpha$ noise characteristic, where $\alpha \approx 1.5$.
At higher drives, the velocity distribution becomes Gaussian and the
fluctuations have a white noise signature.   
We also show that a clogged to unclogged transition
can be produced when the obstacle lattice is diluted through the random
removal of a fraction of obstacles.
The disk arrangement becomes increasingly heterogeneous for increasing
dilution, and a transition to an unclogged state
occurs for dilution fractions close to $0.5$,
indicating that the transition has a percolative character. Our results
should be relevant for clogging dynamics in soft colloidal systems,
emulsions, and bubbles. Similar clogging
effects could also occur for magnetic bubbles,
skyrmions, or superconducting vortices moving
though periodic pinning or obstacle arrays.       

\section*{Conflicts of interest}
There are no conflicts to declare.

\section*{Acknowledgements}
This work was supported by the US Department of Energy through
the Los Alamos National Laboratory.  Los Alamos National Laboratory is
operated by Triad National Security, LLC, for the National Nuclear Security
Administration of the U. S. Department of Energy (Contract No.~892333218NCA000001).



\balance


\bibliography{mybib} 
\bibliographystyle{rsc} 

\end{document}